\documentclass[10pt,twocolumn,aps,prd,superscriptaddress,nofootinbib]{revtex4-2}

\usepackage[T1]{fontenc}
\usepackage[svgnames]{xcolor}
\usepackage[colorlinks,linkcolor=Red,citecolor=blue,urlcolor=blue]{hyperref}

\usepackage{aas_macros_v61}
\usepackage{bm}
\usepackage{orcidlink}
\usepackage{graphicx}
\usepackage{enumerate}

\usepackage{amsmath, amssymb}
\usepackage{ragged2e}

\usepackage{xspace}
\def\Sref#1{Sec.~\ref{#1}\xspace}
\def\Fref#1{Fig.~\ref{#1}\xspace}
\def\Eref#1{Eq.~\eqref{#1}\xspace}
\def\Tref#1{Table~\ref{#1}\xspace}

\begin{document}

\title{Gravitational lensing of fast radio bursts:\\ prospects for probing microlens populations in lensing galaxies}

\author{Ashish Kumar Meena\orcidlink{0000-0002-7876-4321}}
\email{ashishmeena766@gmail.com}
\affiliation{Department of Physics, Indian Institute of Science, Bengaluru 560012, India}
\affiliation{Physics Department, Ben-Gurion University of the Negev, PO Box 653, Be'er-Sheva 84105, Israel}

\author{Prasenjit Saha\orcidlink{0000-0003-0136-2153}}
\affiliation{Physik-Institut, University of Zurich, Winterthurerstrasse 190, 8057 Zurich, Switzerland}

\date{\today}

\begin{abstract}
Gravitational lensing by a stellar microlens of mass~$M$ forms two images separated by micro-arcseconds on the sky and has a time delay of~$2\times10^{-5}(M/{\rm M_\odot})$~seconds. Although we cannot resolve such micro-images in the sky, they could be resolved in time if the source is a fast radio burst~(FRB). In this work, we study the magnification~($|\mu|$) and time delay~($t_d$) distributions of micro-images led by different microlens populations. We find that, in microlensing of typical strongly lensed (macro-)images in galaxy lenses, micro-images stemmed from a population of stellar mass microlenses in the~$[0.08, 1.5]\:{\rm M_\odot}$ range and a second (dark) microlens population in~$[10^{-3} - 10^{-2}]\:{\rm M_\odot}$ range reside in different parts of~$|\mu|-t_d$ plane. For the global minimum macro-image, due to low stellar mass density, we find that the stellar population leads to peaks in autocorrelation at~${>}10^{-6}$~seconds, whereas the secondary population leads to peaks at~${<}10^{-6}$~seconds, allowing us to differentiate different microlens populations. However, an increase in stellar density introduces new peaks at~${<}10^{-6}$~seconds, which can pollute the inference about the presence of multiple microlens populations. In addition, we also show that the number of micro-images, hence the number of peaks in the autocorrelation, is sensitive to the underlying stellar mass function, allowing us to constrain the stellar initial mass function~(IMF) with FRB microlesning in the future. This work is a first step towards using FRB lensing to probe the microlens population within strong lenses, and more detailed studies are required to assess the effect of various uncertainties that we only discussed qualitatively. 
\end{abstract}

\maketitle

\section{Introduction}
\label{sec:intro}
\textit{Gravitational lensing} refers to the bending/deflection of light rays from a background source as they pass close to an intermediate (lens) mass distribution~\citep[e.g.,][]{1986ApJ...310..568B, 1992grle.book.....S}. Depending on the lens mass distribution and the alignment of the lens and source on the sky, gravitational lensing can produce multiple (de-)magnified lensed images of the background source, a regime known as \textit{strong lensing}. The angular separation between these different lensed images is~$\mathcal{O}(10^{-6}\:'')$ for stellar mass lenses, $\mathcal{O}(1'')$ for galaxy lenses, and $\mathcal{O}(1')$~for galaxy cluster lenses. The different path lengths and potentials for various lensed images lead to time delays between these images, which is~$\mathcal{O}(10^{-5})$~seconds for stellar mass lenses, $\mathcal{O}({\rm months})$~for galaxy-scale lenses, and $\mathcal{O}({\rm years})$~for galaxy-cluster lenses. Observations of these time delays can be (and have been) utilized to constrain various cosmological parameters, especially the Hubble constant~\citep[e.g.,][]{2020MNRAS.498.1420W, 2023Sci...380.1322K, 2025ApJ...979...13P}.

So far, we have measured time delays on scales of days to years between multiple lensed images of quasars and supernovae in galaxy and cluster lenses~\citep[e.g.,][]{2018A&A...616A.183B, 2019A&A...629A..97B, 2021NatAs...5.1118R, 2022ApJ...937...34M, 2023ApJ...948...93K}\footnote{we also note the claimed observations of time delays in~$\mathcal{O}([10^{-3}, 1])$~second range presented in~\citet{2021NatAs...5..560P} and~\citet{2025MNRAS.537L..61C} for a lensed gamma-ray burst and fast-radio burst, respectively.}. Every galaxy and cluster lens also contains stellar mass objects~(such as stars, stellar remnants, and planets), which further lens each of the strongly lensed images. Lensing by such objects is known as \textit{microlensing}, as the resulting image separation is~$\mathcal{O}(10^{-6}\:'')$, assuming that the lens is at a cosmological distance. We note that even looking at our solar system, planets can have masses in~$\sim[10^{-6}, 10^{-3}]\:{\rm M_\odot}$ range. Hence, it is more appropriate to call lensing by such objects pico/nano-lensing based on the resulting image separations. However, for the sake of brevity, we also refer to them as microlenses in this work. The time delay between the micro-images is~$\sim2\times10^{-5}\left(M/{\rm M_\odot}\right)$ with~$M$ being the mass of the microlens. Since it is hard to spatially and temporally resolve such micro-images, we typically neglect individual micro-images in the time delay studies of strongly lensed quasars or supernovae. However, the same cannot be said about microlensing-induced variability in the observed brightness and has been studied in detail for both strongly lensed quasars~\citep[e.g.,][]{1998MNRAS.295..573L, 2001PASA...18..207W, 2019ApJ...885...75J, 2024AJ....167..171M} and supernovae~\citep[e.g.,][]{2006ApJ...653.1391D, 2018MNRAS.478.5081F, 2018ApJ...855...22G, 2024SSRv..220...13S}. 

Fast radio bursts~\citep[FRBs;][]{2007Sci...318..777L} are extragalactic microsecond to millisecond-long transient signals observed at radio frequencies in $\sim[0.4, 8.0]$~GHz range with an estimated rate of~$\sim 10^3 - 10^4\:{\rm sky^{-1}\:day^{-1}}$~\citep[e.g.,][]{2019ARA&A..57..417C, 2019A&ARv..27....4P}. Roughly,~2\% to 3\% of these are repeating in nature~\citep{2021ApJS..257...59C, 2023ApJ...947...83C}, although it is claimed that it is only a lower limit and the actual repeating FRB fraction is likely to exceed 50\%~\cite{2024MNRAS.52711158Y}. So far, no strong- or micro-lensed FRB has been fully confirmed, but it is not far in the future, given the large event rate. Thanks to the FRB signal duration of milliseconds, by autocorrelating its intensity, we can constrain the fraction of compact objects, with masses~$\gtrsim10\:{\rm M_\odot}$, in the Universe~\citep[e.g.,][]{2016PhRvL.117i1301M, 2020ApJ...900..122S}, which can be further extended to~$\gtrsim2\:{\rm M_\odot}$ by detecting microlensing of \emph{mini-bursts} within the main FRB burst~\cite{PhysRevD.102.023016}. It may even be possible to achieve nanosecond accuracy in time delay measurements in strongly lensed repeating FRBs and measuring cosmic redshift drift~\cite{2021A&A...645A..44W} by autocorrelating the electric field~(i.e., amplitude of the FRB signal) instead of intensity. With such an excellent time resolution, we can also constrain the abundance of compact dark objects down to masses of~$\simeq10^{-4}~{\rm M_\odot}$~\citep[e.g.,][]{2022PhRvD.106d3016K, 2022PhRvD.106d3017L}. A further possibility is to use micro-images to determine whether the underlying FRB is strongly lensed~\cite{2025arXiv250410523S}. Observing individual micro-images may even allow us to determine the type~(or parity) of the strongly lensed images~\cite{2020MNRAS.497.1583L}.

In lensed quasars (or supernovae), we observe flux variations due to the population of microlenses, i.e., collective microlensing, and it appears that the corresponding magnification distribution is independent of the slope of the microlensing mass function~\citep[e.g.,][]{1992ApJ...386...19W, 1995MNRAS.276..103L, 2001MNRAS.320...21W}. Even if we have a mix of two microlens populations with an order of magnitude difference in their masses, the corresponding magnification distribution can be well approximated by a single microlens population with mass equal to their geometric mean~\citep[e.g.,][]{2020ApJ...904..176E}. Unlike lensed quasars, for microlensed FRBs, we will observe peaks in the autocorrelation that are sensitive to the properties of individual micro-images. This motivates us to ask the following question: Can we discriminate different microlens populations with lensed FRBs, given that we have microsecond~(or higher) temporal resolution in the autocorrelation? In our current work, we take on the above question by simulating different microlensing scenarios for strongly lensed FRBs. Since the separation between peaks and their amplitude in the autocorrelation will be determined by the micro-image time delays and magnifications, respectively~\citep[e.g.,][]{2025arXiv250410523S}, we mainly focus on the magnification and time delay distribution of micro-images. We consider a mix of two microlens populations -- one corresponding to stellar mass microlenses and another representing less massive dark compact objects -- to understand how different microlens populations affect the resulting time delays and magnification distributions and to determine the presence of features in the autocorrelation that can be attributed to the presence of these different microlens populations. In addition, we also study the possibility of distinguishing different stellar initial mass functions~(IMFs) with lensed FRBs.

The current work is organized as follows. In \Sref{sec:basic}, we briefly review the relevant basics of gravitational lensing. In \Sref{sec:pml} and \Sref{sec:pml_pop}, we study the variation of micro-image magnifications as a function of their time delay for an isolated point mass lens and a population of point mass lenses, respectively. In \Sref{sec:sl_ml_pop}, we extend our analysis of micro-image magnifications and their time delays to a mock galaxy-scale strong lens system. \Sref{sec:prob_imf} looks at the effect of variation of the stellar IMF on the magnifications and time delays of the micro-images in strongly lensed images. In \Sref{sec:high_tau}, we study the effect of high microlensing optical depth on the inferences made in the above analyses. In \Sref{sec:limitations}, we discuss the possible origins of uncertainties/limitations of our results in the context of FRBs. We conclude our work in \Sref{sec:conclusions}. Throughout this work, for simplicity, we fix the lens and source redshifts to be 0.5 and 1.5, respectively. We use a flat $\Lambda$CDM cosmology with parameters,~$H_0=70\:{\rm km \: s^{-1} \: Mpc^{-1}}$, $\Omega_{m,0}=0.3$, and $\Omega_\Lambda=0.7$.

\section{Basic lensing}
\label{sec:basic}
A fundamental quantity in gravitational lensing is the arrival time delay surface, which is, up to an additive constant, given as~\citep{1986ApJ...310..568B, 1992grle.book.....S},
\begin{equation}
    t_d(\pmb{x}, \pmb{y}) = \frac{1+z_d}{\rm c} \frac{D_d\:D_s}{D_{ds}} \theta_0^2 \left[ \frac{(\pmb{x} - \pmb{y})^2}{2} - \psi(\pmb{x}) \right],
    \label{eq:lens_td}
\end{equation}
where~$z_d$ is the lens redshift. $D_d$, $D_{ds}$, and $D_s$ are angular diameter distances from observer to lens, lens to source, and observer to source, respectively. $\pmb{y}\equiv\pmb{\beta}/\theta_0$ and~$\pmb{x}\equiv\pmb{\theta}/\theta_0$ represent the dimensionless source position and image plane coordinates, respectively, with~$\theta_0$ being the normalizing angular scale.~$\psi(\pmb{x})$ denotes the (scaled) projected lensing potential. Using Fermat's principle, for a given source position, the image positions are stationary points of the arrival time delay surface~i.e.,~$\nabla_{\pmb{x}}\:t_d(\pmb{x}, \pmb{y}) = 0$. With that, for a given lens model, the relation between the unlensed source position,~$\pmb{y}$, and corresponding lensed image position at~$\pmb{x}$, can be written as,
\begin{equation}
    \pmb{y} = \pmb{x} - \nabla\psi(\pmb{x}),
    \label{eq:lens_eq}
\end{equation}
which is known as the gravitational \textit{lens equation}. Various properties of lensed images can be described by the corresponding Jacobian matrix,
\begin{equation}
    \mathbb{A}(\pmb{x}) \equiv \frac{\partial \pmb{y}}{\partial \pmb{x}} = \delta_{ij} - \psi_{ij},
\end{equation}
where subscript~$i$ and~$j$ represent partial derivatives with respect to image plane coordinates, i.e.,~$\psi_{ij}=\partial^2\psi/\partial x_i \partial x_j$. The magnification,~$\mu(\pmb{x})$, of a lensed image is given as,
\begin{equation}
    \frac{1}{\mu(\pmb{x})} \equiv \det \mathbb{A} = (1-\kappa)^2 - \gamma^2,
\end{equation}
where~$\kappa$ and~$\gamma$ represent the well-known convergence and shear at a given image position in the image plane, respectively. Based on the sign of the eigenvalues of~$\mathbb{A}_{ij}$, lensed images can be divided into three types: minima, maxima, and saddle points. These image types correspond to the nature of the stationary point on the arrival time delay surface.

For a coherent source, such as FRBs, different lensed images will interfere with each other. Thanks to the very high frequencies of FRB signals, we can study the resulting interference under the eikonal approximation. The amplification factor,~$F(f)$, representing the change in signal amplitude and phase due to interference at a frequency,~$f$, can be written as~\cite{1992grle.book.....S, 1998PhRvL..80.1138N, 1999PThPS.133..137N},
\begin{equation}
    F(f) = \sum_{k} \sqrt{|\mu_k|} \exp(2\pi \iota f t_{d,k} - \iota \pi n_k),
    \label{eq:Ff}
\end{equation}
where~$|\mu_k|$ and~$t_{d,k}$ are the magnification and time delay values for the~$k$-th lensed image. $n_k$ is the Morse index with values 0, 1/2, and 1 for minimum, saddle point, and maximum images, respectively. As we can see from the above equation, for typical lensing by galaxy or cluster lenses, interference is not relevant as the time delay between multiple macro-images is too large compared to the signal length. However, it becomes important when studying microlensing as the time delay between micro-images is less than the signal length. Let us assume that we observe a lensed FRB signal in a specific frequency bandwidth, given as,
\begin{equation}
    S_L = \sum_f F(f)\:S(f) = \sum_f F(f) A(f) \exp(2\pi \iota f t),
    \label{eq:lensed_signal}
\end{equation}
where~$S(f) = A(f) \exp(2\pi \iota f t)$ is the unlensed signal at a frequency~$f$. The corresponding autocorrelation function is given as,
\begin{multline}
    C(t) = \int dt' \: S_L(t') \: S_L^\star (t' - t) \\
         = \sum_f \sum_{j,k} |A(f)|^2 \sqrt{|\mu_j \mu_k|}  \exp[-\iota \pi (n_j-n_k)] \\
     \exp[2\pi\iota f (t+t_{d,j}-t_{d,k})].
     \label{eq:auto}
\end{multline}
We can see that lensing introduces peaks in the autocorrelation at~$t=t_{d,k}-t_{d,j}$, with the amplitude of the peak being directly proportional to the geometric mean of micro-image magnifications, i.e.,~$\sqrt{|\mu_j \mu_k|}$. In addition, if the unlensed signal itself contains multiple peaks that are not coherent, they will not introduce additional peaks in the voltage autocorrelation, as they introduce random phase and will not be accumulated in a peak. In contrast, in such a scenario, intensity autocorrelation can have multiple peaks even in the absence of lensing.

An interesting point to note is that since the peak magnitudes in the autocorrelation depend on the magnifications of both~$j$-th and $k$-th images, such that a low magnification in one image can be compensated by a higher magnification in the other. This also implies that if our brightest image has ~$|\mu|\simeq1$, assuming signal-to-noise~(SNR) of~$20-100$, we can expect to detect peaks of micro-images with~$|\mu|\simeq 10^{-1} - 10^{-2}$ above $5\sigma$. With the above, we can argue that if different microlensing populations lead to micro-images with distinct magnification and time delay values, they can be segregated by searching for corresponding features (i.e., peaks) in the resulting autocorrelation. Hence, in our current work, we will primarily focus on~$|\mu|-t_d$ plane and peak distribution in the autocorrelation for various lensing scenarios.

\begin{figure}
    \centering
    \includegraphics[height=6.0cm, width=8.5cm]{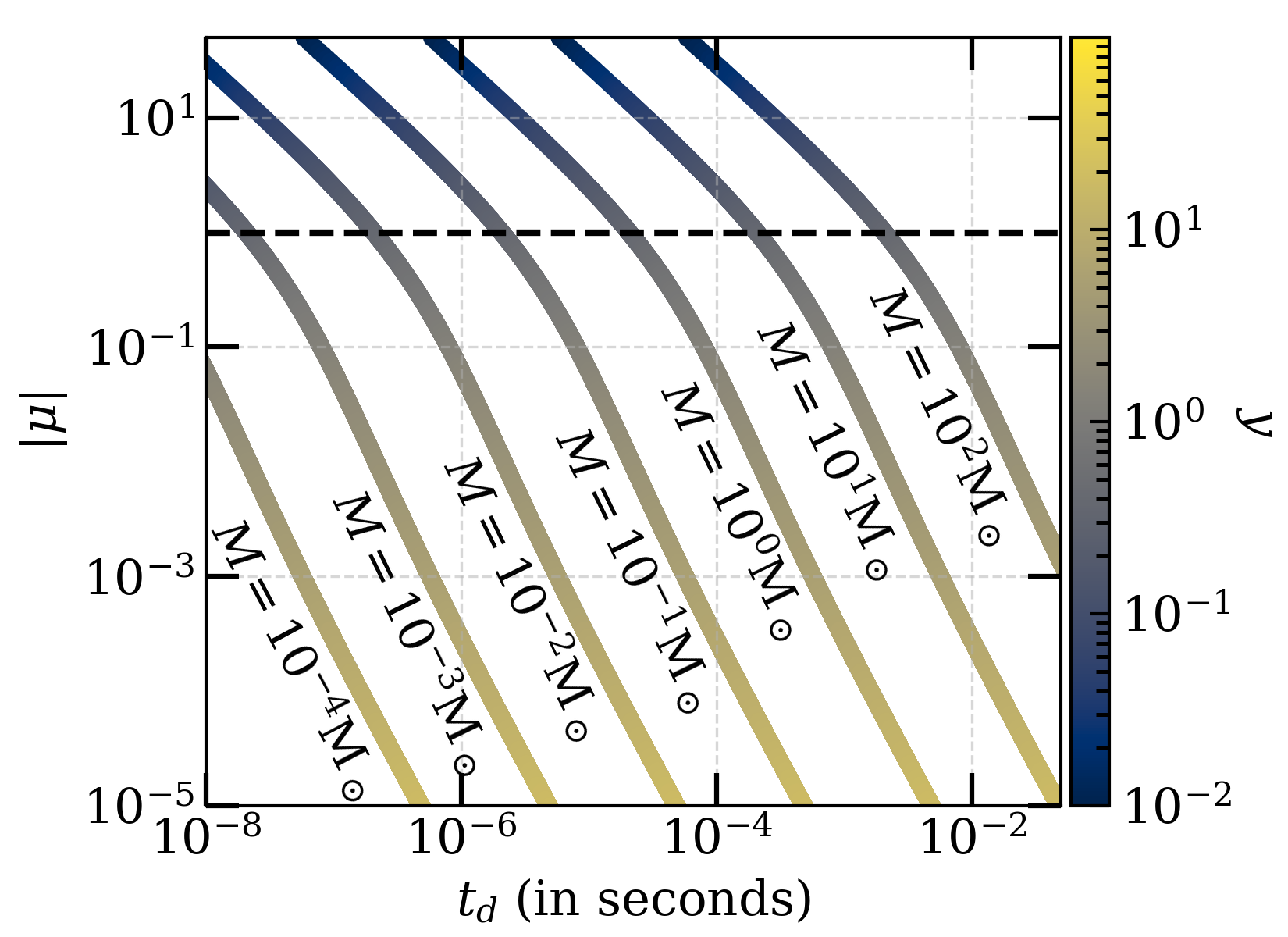}
    \caption{Time delay~($t_d$) vs. absolute magnification~($|\mu|$) for the saddle point image formed by an isolated point mass lens. The different curves correspond to different mass values, which are shown below each curve. Each curve is color-coded according to the source position~($y$). The horizontal dashed black line corresponds to~$|\mu|=1$.}
    \label{fig:pm_mutd}
\end{figure}

\section{point mass lens}
\label{sec:pml}
Arguably, the simplest lens model is an isolated point mass lens (or Schwarzschild lens), for which the lensing potential (in dimensionless form) is given as~\cite{1992grle.book.....S},
\begin{equation}
    \psi(x) = \ln(x),
    \label{eq:pm_psi}
\end{equation}
where the normalizing angular scale~($\theta_0$) is assumed to be equal to the Einstein angle~($\theta_{\rm E}$) corresponding to the lens mass,~$M_s$, given as,
\begin{equation}
    \theta_0 \equiv \theta_{\rm E} = \sqrt{\frac{4{\rm G}M_s}{\rm c^2} \frac{D_{ds}}{D_d\:D_s}}.
    \label{eq:pm_tE}
\end{equation}
A point mass lens always leads to the formation of two images, one minimum and one saddle point. Using lens equation, \Eref{eq:lens_eq}, the image positions are,
\begin{equation}
    x_{\pm} = \frac{y}{2} \pm \frac{\sqrt{y^2+4}}{2},
    \label{eq:pm_impos}
\end{equation}
where `$+$' and `$-$' represent the minimum and saddle point images, respectively. The corresponding magnifications are given as,
\begin{equation}
    \mu_\pm = \frac{1}{2} \pm \frac{y^2 + 2}{2y\sqrt{y^2+4}}.
    \label{eq:pm_mu}
\end{equation}
The time delay of the saddle point image with respect to the minimum image is given as,
\begin{multline}
    t_d(x, y) = \frac{1+z_d}{\rm c} \frac{4 {\rm G} M_s}{\rm c^2} \left[ \frac{y\sqrt{y^2+4}}{2} + \right. \\
            \left. \ln\left( \frac{\sqrt{y^2+4} + y}{\sqrt{y^2+4} - y}  \right) \right].
    \label{eq:pm_td}
\end{multline}

In \Fref{fig:pm_mutd}, we plot the saddle point magnification~($|\mu|$) vs. its time delay~($t_d$) with respect to the minimum as a function of source position~($y$), assuming different point lens mass values in the~$[10^{-4}, 10^2]~{\rm M_\odot}$ range. As we can see from the time delay expression, \Eref{eq:pm_td}, all curves are the same except for a horizontal shift depending on the lens mass, which stems from the overall scaling of time delay with the lens mass. Thanks to this scaling, a more massive point mass lens can lead to bright~($|\mu|\geq1$) saddle point images at large time delays. For example, a point mass lens with~$M_s = 1~{\rm M_\odot}$ can lead to saddle point images with~$|\mu|\gtrsim1$ at~$t_d\lesssim0.02$~milliseconds. On the other hand, a point mass lens with~$M_s = 10^2~{\rm M_\odot}$ can lead to similarly magnified saddle point images up to~$t_d\lesssim2$~milliseconds.

Although the above analysis is very straightforward, it has strong implications. For an isolated point mass lens, we will see a secondary peak in the autocorrelation corresponding to the saddle point. The time at which the secondary peak occurs and its height can be used to determine the properties of the point mass lens~\citep[e.g.,][]{2016PhRvL.117i1301M}. If we have more than one point mass lenses sufficiently far away from each other~(i.e., their critical curves do not overlap with each other), each of the point mass lenses will lead to one saddle point image whose~$|\mu|$ and $t_d$ follow a certain curve in \Fref{fig:pm_mutd} according to its mass. Hence, based on the masses of point lenses, corresponding images are expected to occupy different parts of the~$|\mu|-t_d$ plane, an effect that could be used to distinguish multiple point mass lens populations, given that we can temporally resolve these images.

\begin{figure*}
    \centering
    \includegraphics[height=6.5cm, width=16cm]{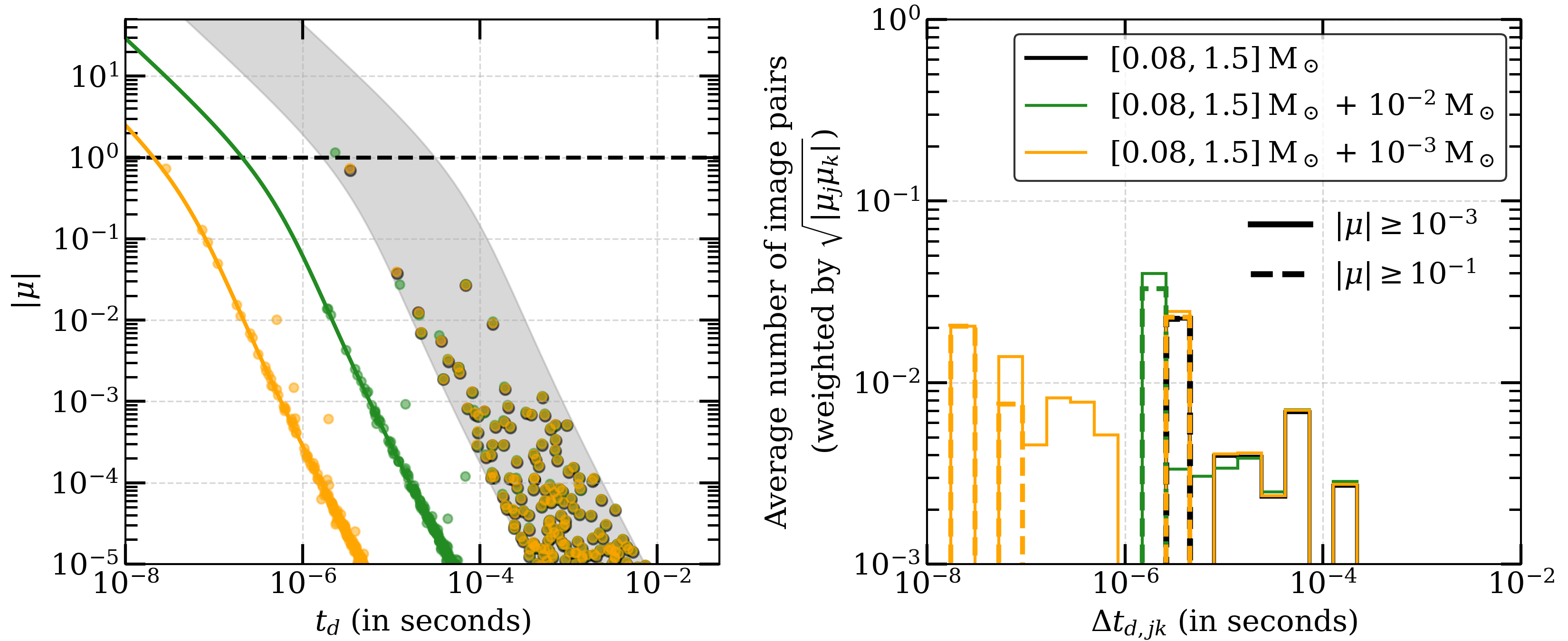}
    \caption{\textit{Left panel.} Time delay~($t_d$) vs. absolute magnification~($|\mu|$) distribution of micro-images corresponding to mock populations of microlenses in the absence of external effects. The black points (which are covered up by other colored points) correspond to a stellar microlens population drawn using the Salpeter mass function assuming~$\tau=10^{-2}$. The gray shaded region covers a range of~$|\mu|-t_d$ values for an isolated microlens mass range of~$[0.08, 1.5]~{\rm M_\odot}$. To generate green~(yellow) points, we have added another microlens population with a constant mass of~$10^{-2}~{\rm M_\odot}$~(~$10^{-3}~{\rm M_\odot})$ with~$\tau=10^{-2}$. The green and yellow solid curves represent the $|\mu|-t_d$ values for an isolated microlens with~$M=10^{-2}~{\rm M_\odot}$ and~$M=10^{-3}~{\rm M_\odot}$, respectively. The horizontal black dashed line corresponds to~$|\mu|=1$. \textit{Right panel.} Average histogram of time delays~($\Delta t_{d,jk}$) between pairs of micro-images shown in the left panel, weighted by the corresponding geometric mean of magnifications, i.e.,~$\sqrt{|\mu_j \mu_k|}$. The solid and dashed histograms are made using micro-images with~$|\mu|\geq10^{-3}$ and~$|\mu|\geq10^{-1}$, respectively.}
    \label{fig:pm_population}
\end{figure*}

\section{Population of point mass lenses}
\label{sec:pml_pop}
Instead of an isolated point mass lens, if we have collective microlensing by a population of~$N$ point mass lenses, i.e., a random star field without external effects, the corresponding lensing potential at a position~$\pmb{x}$ can be written as a linear superposition of individual potentials scaled by the corresponding masses,
\begin{equation}
    \psi(\pmb{x}) = \sum_{i=1}^N \frac{M_i}{M_s} \ln (|\pmb{x} - \pmb{x}_i|),
\end{equation}
where~$M_i$ represents the mass of $i$-th microlens and~$M_s$ is an arbitrary normalizing mass scale. Other quantities, such as time delays and magnifications, can be derived using the formulae in \Sref{sec:basic}. Since we do not have any external effects, without microlenses (i.e., no lensing), the arrival time delay surface would have had a parabolic shape with one minimum. Hence, the current setup can be considered equivalent to microlensing of a minimum image in the outskirts of a lensing galaxy, where the effects due to the overall galaxy are negligible.

To determine the number of microlenses, we use the concept of optical depth, which is defined as the fraction of area~(on the sky) covered by the critical curves of point mass lenses and can be written as,
\begin{equation}
    \tau \equiv \frac{N \pi \langle\theta_E\rangle^2}{\rm Area},
    \label{eq:tau_ml}
\end{equation}
where~$\langle\theta_E\rangle$ is the average Einstein angle corresponding to the average mass,~$\langle M \rangle$, of the microlens population. For example, the Salpeter mass function~\cite{1955ApJ...121..161S} in the~$[0.08, 1.5]~{\rm M_\odot}$ mass range has an average mass of~$\langle M \rangle = 0.2~{\rm M_\odot}$. To study the~$|\mu|-t_d$ distribution of micro-images corresponding to different microlens populations, here, we restrict ourselves to low optical depth values and simulate three cases:
\begin{enumerate}[(i)]
    \item stellar microlens population with~$\tau=10^{-2}$ and drawn using the Salpeter mass function in the~$[0.08, 1.5]~{\rm M_\odot}$ mass range,
    \item stellar microlens population in case~(i) and a second microlens population with~$\tau=10^{-2}$ with a constant microlens mass of~$10^{-2}~{\rm M_\odot}$,
    \item stellar microlens population in case~(i) and a second microlens population with~$\tau=10^{-2}$ with a constant microlens mass of~$10^{-3}~{\rm M_\odot}$.
\end{enumerate}
For each of the above cases, we simulate 50~realizations. Here, we note that~$\tau=10^{-2}$ is equivalent to a microlens density of~$23.284\:{\rm M_\odot \: pc^{-2}}$, which is similar to stellar densities in the outskirts of lensing galaxies. We refer to the microlens population in~$[0.08, 1.5]~{\rm M_\odot}$ mass range as the stellar population, as we are focusing on galaxy-scale lenses which are typically early-type galaxies~(ETGs) and have little to no star formation. Hence, such galaxies are expected to harbor stellar populations older than $\sim2-3$~Gyr, a time in which all stars with mass~$\gtrsim1.5~{\rm M_\odot}$ are expected to complete their life. We note that the chosen values for mass range and fraction of secondary microlens population are somewhat ad hoc, but they fall within the allowed parameter space and are not fully excluded by current constraints from local microlensing surveys~\citep[e.g.,][]{2001ApJ...550L.169A, 2007A&A...469..387T, 2024Natur.632..749M}.

For microlensing simulations, we assume that the source is placed at the origin in the source plane. Next, we distribute microlenses in a circular region in the image plane whose size is determined by the tangential and radial macro-magnification values while ensuring that the contribution from outside this region is negligible. After that, we divide the image plane box into~$1000\times1000$ pixels, and in the next step, each pixel is further divided into~$64\times64$ pixels. We find crossings of contour corresponding to $\beta_x - \theta_x + \alpha_x = 0$ and $\beta_y - \theta_y + \alpha_y = 0$, which mark the position of micro-images. Since we primarily focus on typical macro-magnifications in this work, with the final image plane resolution~$<10^{-9}$~arcseconds, we can be confident that we do not expect to miss any micro-image with~$|\mu|\gtrsim10^{-7}$.

The resulting~$|\mu|-t_d$ distributions of micro-images are shown in the left panel of \Fref{fig:pm_population}. Here, we plot micro-image time delays with respect to the global minimum image situated at~$t_d=0$. For case~(i), shown by black points, we note that micro-images formed due to the stellar mass microlenses always lie in the gray band which corresponds to the region of~$|\mu|-t_d$ plane covered by isolated point masses in~$[0.08, 1.5]~{\rm M_\odot}$ mass range~(see \Sref{sec:pml}). This implies that for the low optical depth, $|\mu|-t_d$ values follow the isolated point mass case very well. This can be understood by noting that at low optical depth, properties of a saddle point are primarily determined by the microlens that created it, while the other microlenses are sufficiently far away to have a significant effect. For case~(ii), shown by green points, we note the formation of a secondary branch of micro-images in addition to micro-images in the gray shaded region. These micro-images are a result of the second microlens population with~$M=10^{-2}{\rm M_\odot}$ and they follow closely the green curve, which represents the~$|\mu|-t_d$ curve for an isolated point mass lens with~$M=10^{-2}{\rm M_\odot}$. A similar behavior is also observed for case~(iii), shown by yellow points, except that the secondary branch of micro-images follows the yellow curves, which is for an isolated point mass lens with~$M=10^{-3}{\rm M_\odot}$. Hence, we can infer that micro-images created by different microlens populations in the low optical depth regime lie in distinct regions of~$|\mu|-t_d$ plane. We note that in all of the 50~realizations, except the global minima, nearly all micro-images have~$|\mu|\lesssim1$. This is a result of the fact that in low optical depth, it is less likely to have a microlens aligning well with the source position, so that it can lead to multiple bright~($|\mu|\geq1$) micro-images. In our simulations, for case~(i) the average number of micro-images per realization with~$|\mu|>(10^{-1}, 10^{-2}, 10^{-3})$ is $(1.02, 1.08, 1.28)$ including the global minima at~$t_d=0$. For case~(ii) and case~(iii), looking at \Fref{fig:pm_population}, we do not expect to see a large change in the number of micro-images for~$|\mu|>10^{-2}$ but see a considerable increase in fainter images with for~$|\mu|>10^{-3}$.

Another very interesting point to note is that micro-images primarily form at larger time delays compared to an isolated point mass lens, i.e., micro-images lie rightwards of the lower limit of the gray shaded region, green curve, and yellow curve for case~(i), (ii), and~(iii), respectively\footnote{We note some micro-images slightly leftwards but very close to green/yellow curve with~$|\mu|<10^{-3}$. This slight leftwards shift is likely to be a resolution effect.}. Such a behavior can be understood from the fact that in low optical depth, we would primarily expect to have one minimum and a large number of saddle points for each microlens. In low optical depth, these saddle points are expected to follow the isolated point mass case. That said, sometimes multiple microlenses can lie close to each other such that the resulting saddle-point micro-image can have relatively higher time delay and magnification values, and end up rightward of, for example, the green curve, as we do see for some micro-images. However, this does not mean that we can never have micro-images to the left of the curves predicted by an isolated point mass model. For example, if one of the microlenses aligns well such that the source lies within the corresponding diamond caustic, it will give rise to two minima, and the second minimum can have a smaller time delay value than the saddle points~(as we will also see~\Sref{sec:high_tau} for high optical depth). That said, based on our simulations, this seems to be rare for low microlensing optical depths.

The right panel of \Fref{fig:pm_population} shows the histogram of time delays between pairs of micro-images, with each pair weighted by the geometric mean of the corresponding magnifications. All histograms are averaged over the 50 realizations. Looking at \Eref{eq:auto}, we can see that such a plot represents the distribution of peaks in the autocorrelation. We can see that the stellar population primarily leads to peaks with~$\Delta t_{d,jk}>10^{-6}$, and the presence of a secondary population starts to give rise to peaks at~$\Delta t_{d,jk} \lesssim 10^{-6}$. We note that if we only consider micro-images with~$|\mu|\geq10^{-1}$~(dashed histograms), the peaks in the autocorrelation for multiple branches corresponding to different cases are separated in time. For example, we see peaks in the yellow histogram at~$\sim10^{-5}$~seconds and at~$\sim10^{-7}$~seconds, which originate due to the presence of stellar microlens population and compact objects with masses of~$10^{-3}\:{\rm M_\odot}$, respectively. This implies that we can determine the presence of different microlens populations. However, once we consider fainter images up to~$|\mu|\geq10^{-3}$~(solid histograms), we start to see that peaks in autocorrelation span a wider range in time, especially for the yellow histogram. That said, even now, we do not see peaks below~$\Delta t_{d,jk}<10^{-6}$ for stellar microlenses. This implies that, at least for the low optical depth case~($\tau\simeq10^{-2}$), peaks at~$\Delta t_{d,jk}<10^{-6}$ in the autocorrelation can be a strong evidence for microlensing due to a dark compact object. In addition, comparing the green and yellow histograms reveals that a larger difference in the mass of stellar populations and compact objects improves the chances of identification of multiple microlens populations.

\begin{figure}
    \centering
    \includegraphics[height=6cm, width=8cm]{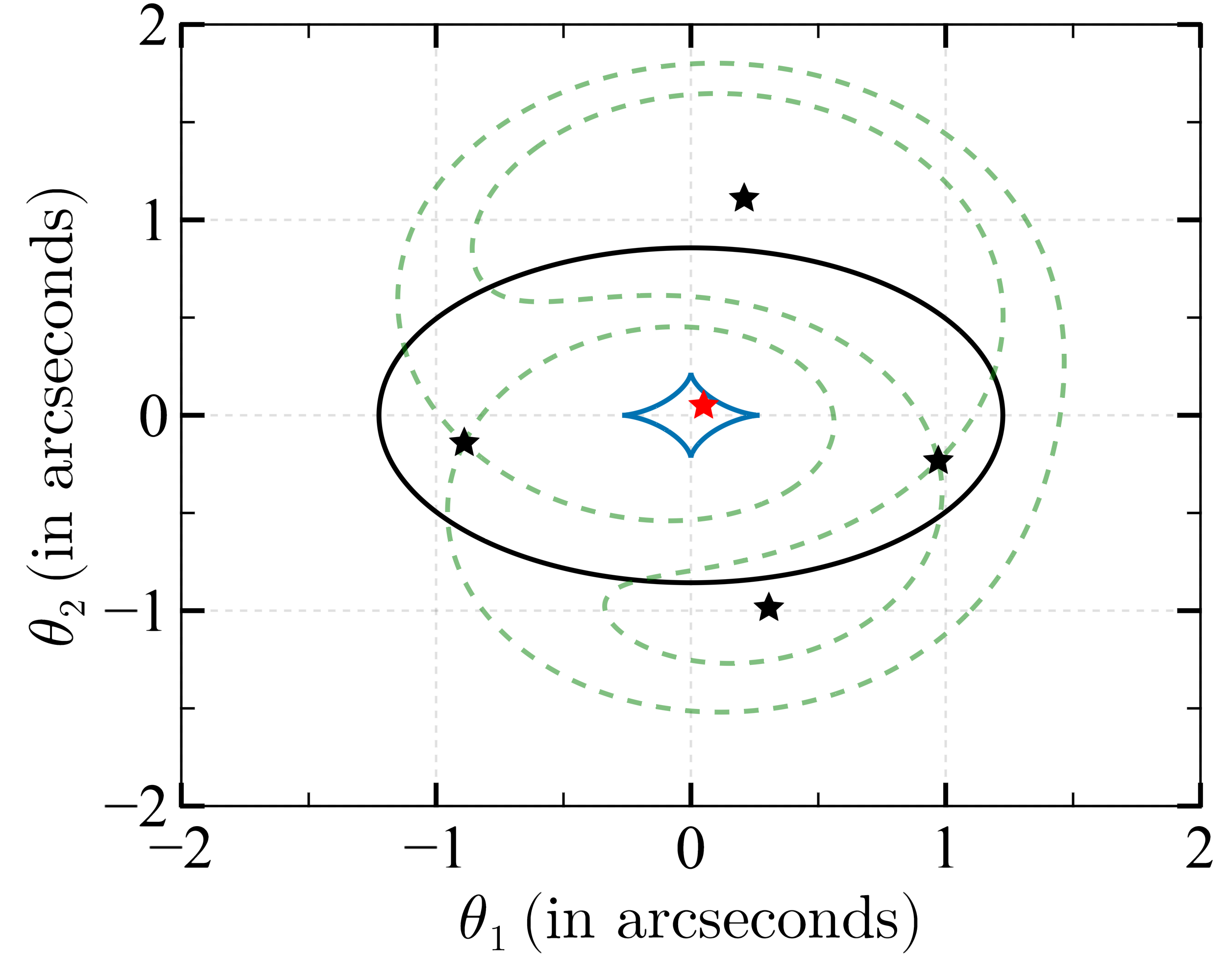}
    \caption{Mock galaxy scale strong lens system assuming SIE mass density profile with~$(\sigma, \epsilon)=(250\:{\rm km\:s^{-1}}, 0.3)$. Black and blue curves represent the critical curve and caustic, respectively. The red star marks the source position, and the black stars show the corresponding image positions. The green dashed curves represent the arrival time delay contours corresponding to saddle point images.}
    \label{fig:macrolens}
\end{figure}

\begin{figure*}
    \centering
    \includegraphics[height=6.5cm, width=16cm]{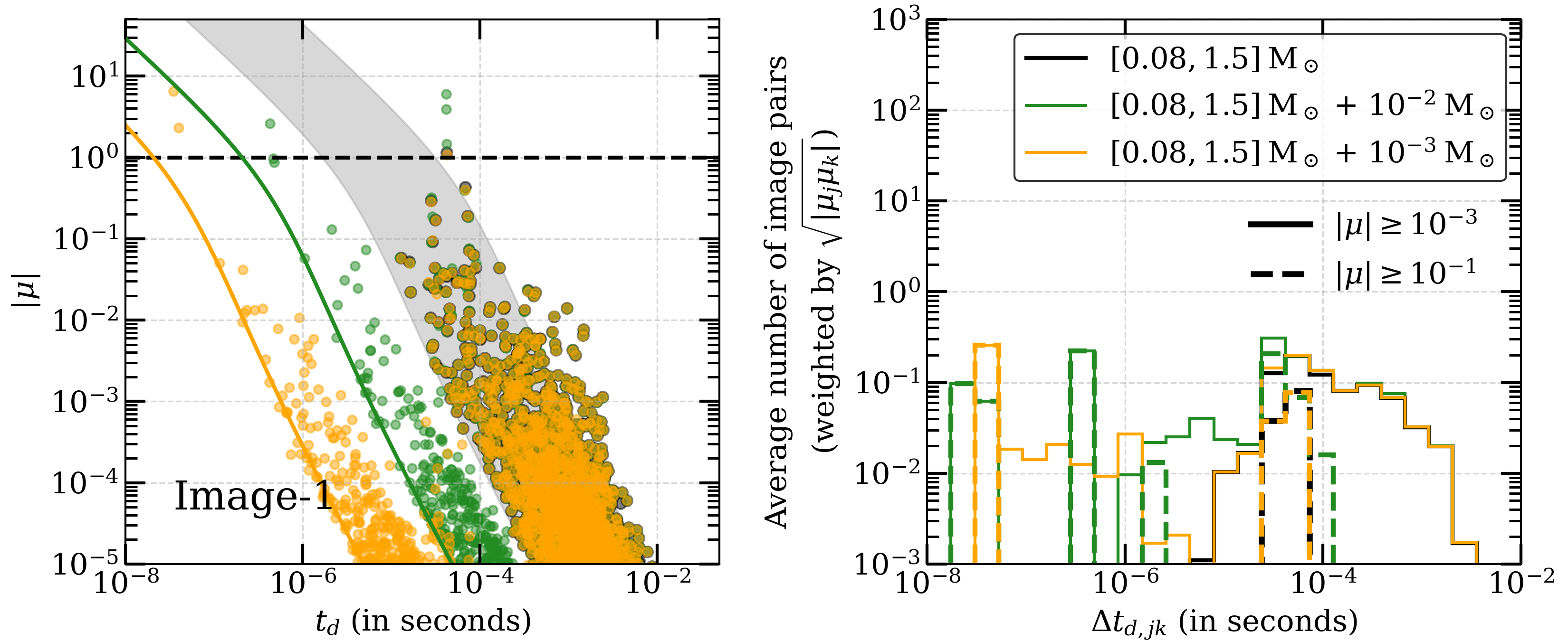}
    \includegraphics[height=6.5cm, width=16cm]{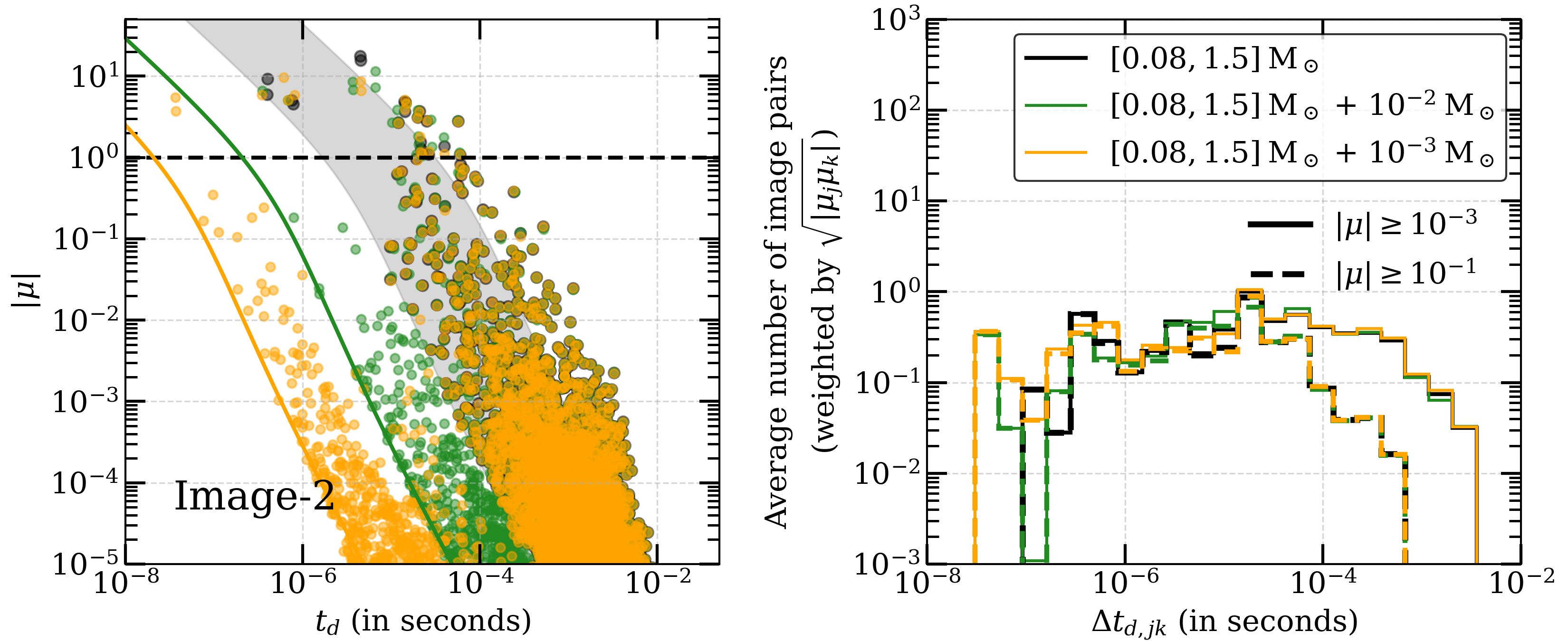}
    \includegraphics[height=6.5cm, width=16cm]{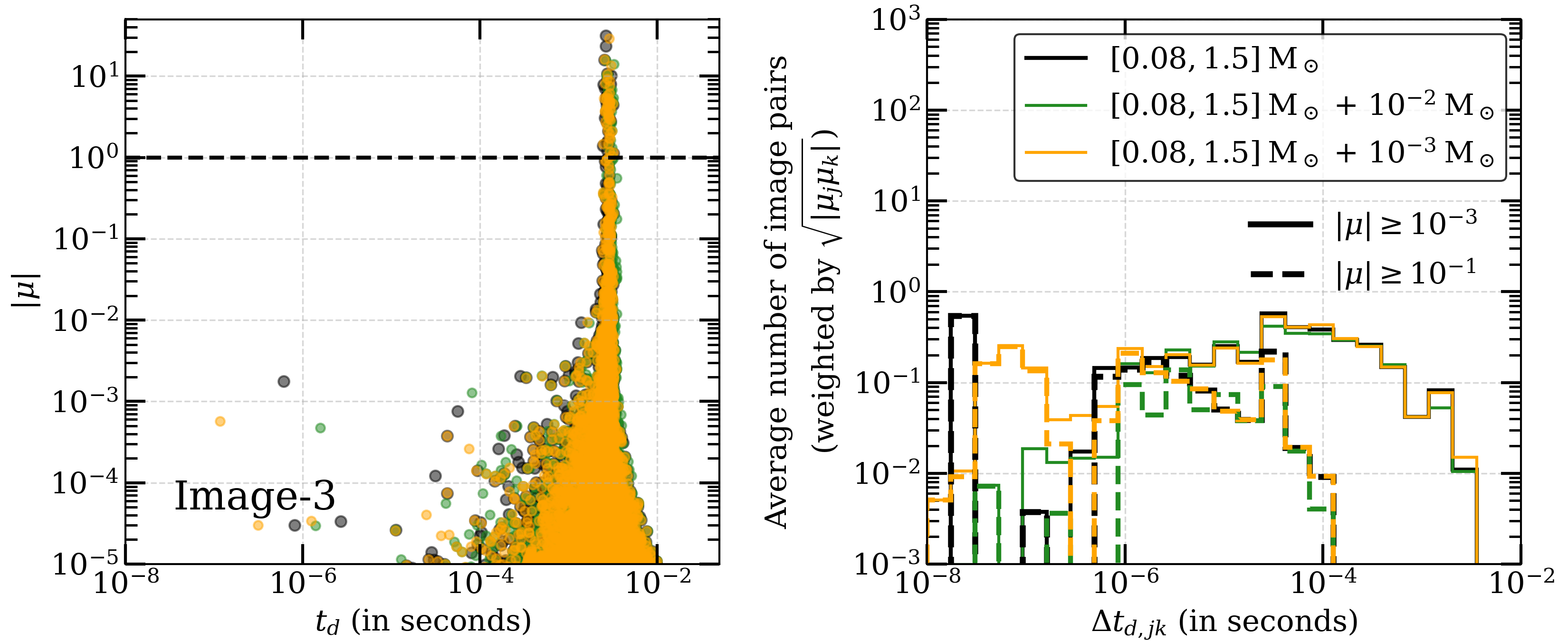}
    \caption{\textit{Left column.} Time delay~($t_d$) vs. absolute magnification~($|\mu|$) distribution of micro-images for the first three macro-images in our mock strong lens system. The macro-images are labelled according to arrival time~(see Table~\ref{tab:sl_system} for more details). In each panel, the black points (which are covered up by other colored points) correspond to a stellar microlens population drawn using the Salpeter mass function with the surface density given in \Tref{tab:sl_system}. To generate green~(yellow) points, for each image, we have added a second microlens population equivalent to 1\% of total convergence of constant mass,~$10^{-2}{\rm M_\odot}$~($10^{-3}~{\rm M_\odot}$). The horizontal black dashed line corresponds to~$|\mu|=1$. \textit{Right column.} Average histogram of time delays~($\Delta t_{d,jk}$) between pairs of micro-images shown in the corresponding left panels, weighted by the corresponding geometric mean of magnifications, i.e.,~$\sqrt{|\mu_j \mu_k|}$. The solid and dashed histograms only include micro-images with~$|\mu|\geq10^{-3}$ and~$|\mu|\geq10^{-1}$, respectively.}
    \label{fig:sl_population}
\end{figure*}

\section{Microlensing of strongly lensed signals}
\label{sec:sl_ml_pop}
To simulate a realistic case of microlensing of strongly lensed (macro-)images, we consider an isolated ETG with velocity dispersion of~$250~{\rm km\:s^{-1}}$ and an ellipticity~($\epsilon$) of 0.3 as a lens represented by a singular isothermal ellipsoid~\citep[SIE;][]{2018pgl..book.....C} density profile. The critical curve and caustic for such a lens are shown in \Fref{fig:macrolens}. We choose a source inside the diamond caustic (red star), leading to four lensed images whose details are given in \Tref{tab:sl_system}. The first two macro-images are minima, and the remaining two are saddle points. The stellar surface density at the position of each macro-image has been calculated assuming a S\'esric profile for light distribution following~\citet{2019MNRAS.483.5583V}.

To study the microlensing of different macro-images, in their vicinity, we can approximate the effect of the overall galaxy as an external effect given by constant convergence and shear values. The resulting lensing potential can be written as,
\begin{multline}
    \psi(\pmb{x}) = \sum_{i=1}^n \frac{M_i}{M_s} \ln (|\pmb{x} - \pmb{x}_i|) + \frac{\kappa_s}{2} (x_1^2 + x_2^2) \\ 
    + \frac{\gamma_1}{2} (x_1^2 - x_2^2) + \gamma_2 \: x_1 \: x_2,
    \label{eq:psi_slml}
\end{multline}
where~$\kappa_s \equiv \kappa - \kappa_{\star}$ is the smooth convergence with~$\kappa$ representing the total convergence and~$\kappa_\star$ being the convergence in the form of microlenses. $(\gamma_1, \gamma_2)$ are the shear components due to the overall galaxy lens. $M_i$ is the mass of the $i$-th microlens and $M_s$ is an arbitrary mass value for scaling. Without loss of generality, locally, we can always choose the principal direction of shear to be aligned with the abscissa~(i.e., $|\gamma| = \sqrt{\gamma_1^2 + \gamma_2^2} = |\gamma_1|$ and $\gamma_2 = 0$). Here again, we simulate three cases for the first three strongly lensed images (i.e., two minima and one saddle point): 
\begin{enumerate}[(i)]
    \item stellar microlens population drawn using the Salpeter mass function in the~$[0.08, 1.5]~{\rm M_\odot}$ mass range with surface density given in \Tref{tab:sl_system},
    \item stellar microlens population in case~(i) and a second microlens population with 1\% of total convergence,~$\kappa$, in form of microlenses with~$M=10^{-2}\:{\rm M_\odot}$,
    \item stellar microlens population in case~(i) and a second microlens population with 1\% of total convergence,~$\kappa$, in form of microlenses with~$M=10^{-3}\:{\rm M_\odot}$.
\end{enumerate}

\begin{table}
    \centering
    \caption{Details of mock galaxy scale lens system shown in~\Fref{fig:macrolens}. $x$ and $y$ represent the macro-image positions for a source at~(0.05, 0.05) arcseconds with respect to the lens center. $\mu$ and $t_d$ represent the macro-magnification and time delay. $\kappa$ and $\gamma$ describe the convergence and shear at the position of each macro-image. $\kappa_\star$ and $\Sigma_\star$ are the stellar convergence and the corresponding stellar density. The last column represents the stellar microlensing optical depth~($\tau_\star$; without including macro-magnification), assuming an average mass of~$\langle M \rangle = 0.2\:{\rm M_\odot}$.}
    \label{tab:sl_system}
    \begin{tabular}{ccccccccc}
     \hline
     $x$    & $y$    & $\mu$ & $t_d$  & $\kappa$ & $\gamma$ & $\kappa_\star$ & $\Sigma_\star$             & $\tau_\star$ \\
     ($''$) & ($''$) &       & (days) &          &          &                & (${\rm M_\odot\:pc^{-2}}$) &              \\
     \hline
     0.21   & 1.11   & 4.4   & 0.0    & 0.383    & 0.383    & 0.028          & 65.0                       & 0.03         \\
     0.31   &-0.99   & 6.7   & 9.8    & 0.425    & 0.425    & 0.036          & 82.8                       & 0.04         \\
     0.97   &-0.23   &-5.2   & 13.0   & 0.596    & 0.596    & 0.075          & 174.0                      & 0.08         \\
    -0.89   &-0.14   &-2.9   & 21.7   & 0.671    & 0.671    & 0.095          & 222.3                      & 0.10         \\
     \hline
    \end{tabular}
\end{table}

We again simulate 50 realizations and the results are shown in \Fref{fig:sl_population}. For macro minimum images (Image-1 and Image-2) in the left column, we again observe the formation of multiple micro-image branches, similar to those in the left panel of \Fref{fig:pm_population}, with some interesting differences. For example, the green points now cover the whole area between the green curve and the gray shaded region. This can be understood from the fact that now the microlenses are embedded in a strong lens, which introduces non-zero macro-magnification. This macro-magnification changes the effective mass of microlenses, which in turn affects the micro-image time delays. This also explains the formation of micro-images rightward of the gray shaded region. That said, even now, it is hard to produce cases where black points lie leftwards of the gray shaded region. The same is also true for green and yellow points.

\begin{figure*}
    \centering
    \includegraphics[height=6.5cm, width=16cm]{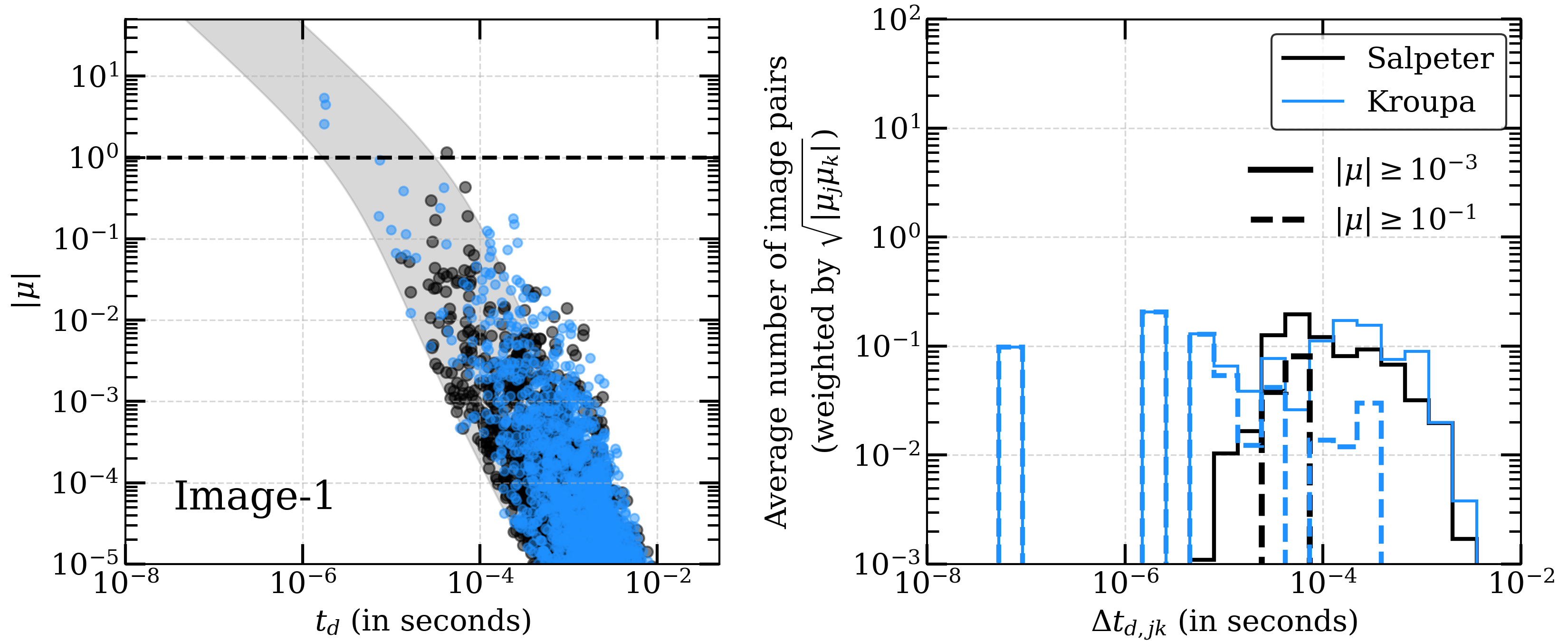}
    \includegraphics[height=6.5cm, width=16cm]{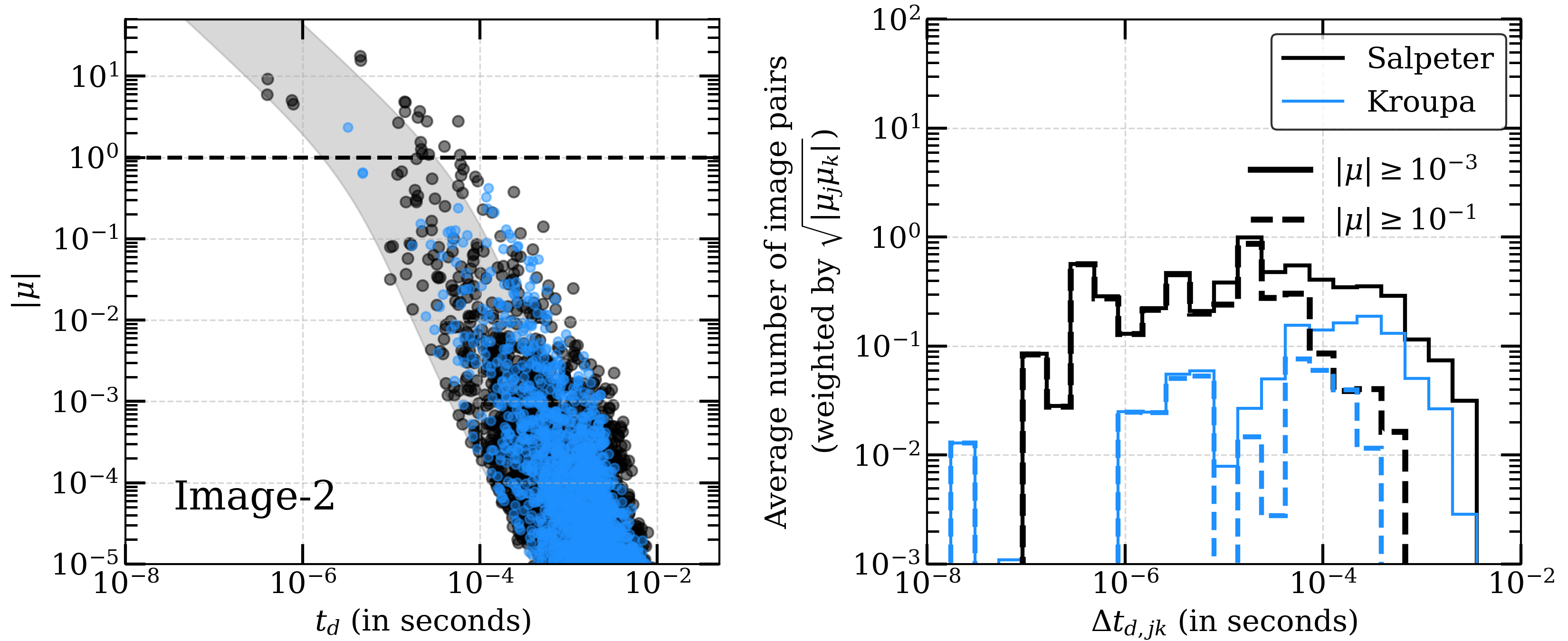}
    \caption{\textit{Left column.} Time delay~($t_d$) vs. absolute magnification~($|\mu|$) distribution of micro-images corresponding to different stellar mass functions for Image-1/2 in \Tref{tab:sl_system}. For each macro-image, in the bottom scatter plot, the black points correspond to the Salpeter mass function, whereas the blue points are for the Kroupa mass function. The gray shaded region covers the range of~$|\mu|-t_d$ values for an isolated microlens mass in~$[0.08, 1.5]~{\rm M_\odot}$ range. The horizontal black dashed line corresponds to~$|\mu|=1$. \textit{Right column.} Average histogram of time delays~($\Delta t_{d,jk}$) between pairs of micro-images shown in the corresponding left panels, weighted by the corresponding geometric mean of magnifications, i.e.,~$\sqrt{|\mu_j \mu_k|}$. The solid and dashed histograms include micro-images with~$|\mu|\geq10^{-3}$ and~$|\mu|\geq10^{-1}$, respectively.}
    \label{fig:pm_sim_imf}
\end{figure*}

For Image-1, in case~(i), the average number of micro-images in each realization with~$|\mu|>(10^{-1}, 10^{-2}, 10^{-3})$ is~$(1.1, 1.92, 4.98)$. For Image-1 (top-left panel), in one of our realizations, we note an interesting case of a stellar micro-image further splitting into three micro-images due to the presence of a compact dark object. At~$(t_d, |\mu|)\sim(10^{-4}, 1)$, we see a black point just outside of the gray shaded region representing a bright micro-image induced due to the presence of stellar microlenses. Close to it, we see three green points aligned vertically and showing the splitting of the one micro-image into three micro-images, which is a result of the chance alignment of a dark compact object with the position of stellar micro-images. For yellow points, the same alignment did not occur, and we again see only a single stellar micro-image. In the top-right panel, we observe that stellar microlenses alone are unable to lead to micro-image pairs with time delay~$\lesssim10^{-6}$, and only the presence of a second microlens population can do so. Comparing with \Fref{fig:pm_population}, we now have pairs of micro-images with~$\sqrt{|\mu_i \mu_j|}\geq0.1$ and time delays~$\simeq10^{-4}$~seconds corresponding to stellar microlenses, which can be attributed to the formation of micro-images at larger time delays due to macro-magnification provided by the overall galaxy lens.

For Image-2 (middle-left panel), we can clearly see an increase in the number of bright micro-images~(i.e.,~$|\mu|\geq1$) compared to Image-1, in the gray shaded region, which is a result of an increase in stellar surface density and the same can be expected for the secondary micro-image population if we increase the corresponding surface density. The average number of micro-images with~$|\mu|>(10^{-1}, 10^{-2}, 10^{-3})$ is ~$(1.96, 3.74, 8.56)$ for case~(i). This increase in the overall number of micro-images results in image pairs (with $\sqrt{|\mu_i \mu_j|}\geq0.1$) covering a wider range of time delays, $\sim[10^{-7}, 10^{-3}]$~seconds, compared to Image-1, as we see in the middle-right panel. This will hinder our capability of detecting the second microlens population, as the stellar microlenses alone can lead to micro-image pairs with such short time delays. That said, we still see micro-image pairs with time delays $\simeq5\times10^{-8}$~seconds in the green/yellow histogram, which corresponds to micro-images formed due to the second microlens population, but again, segregating images at such extremely short time delays can be challenging as we discuss in \Sref{sec:limitations}.

For macro saddle point~(Image-3), the average number of micro-images with $|\mu|>(10^{-1}, 10^{-2}, 10^{-3})$ is~$(1.8, 4.1, 12.68)$ for case~(i).
This increase in the number of bright micro-images is hard to observe in \Fref{fig:sl_population} due to the geometry of the arrival time delay surface around a saddle point, as we observe micro-images forming before and after the brightest micro-image~(also see Fig.~6 in~\cite{2020MNRAS.497.1583L}). As a result, we see micro-images corresponding to different cases occupying the same parts of the~$|\mu|-t_d$ plane, unlike macro-minimum images~(i.e., Image-1 and Image-2). This hinders our ability to make inferences about the underlying microlens mass function directly from~$|\mu|-t_d$. Looking at the micro-image pair time delays, we again observe that stellar micro-images alone can cover a wide range of time delays similar to Image-2, which will make it hard to recognize the presence of multiple microlens populations, at least for the optical depths considered here. Since stellar microlens density is even higher for Image-4, which is a macro saddle point, we expect similar behavior for that also. In addition, considering the increase in computation time due to the higher stellar density, we excluded Image-4 from our current analysis.

\begin{figure*}
    \centering
    \includegraphics[height=6.5cm, width=16cm]{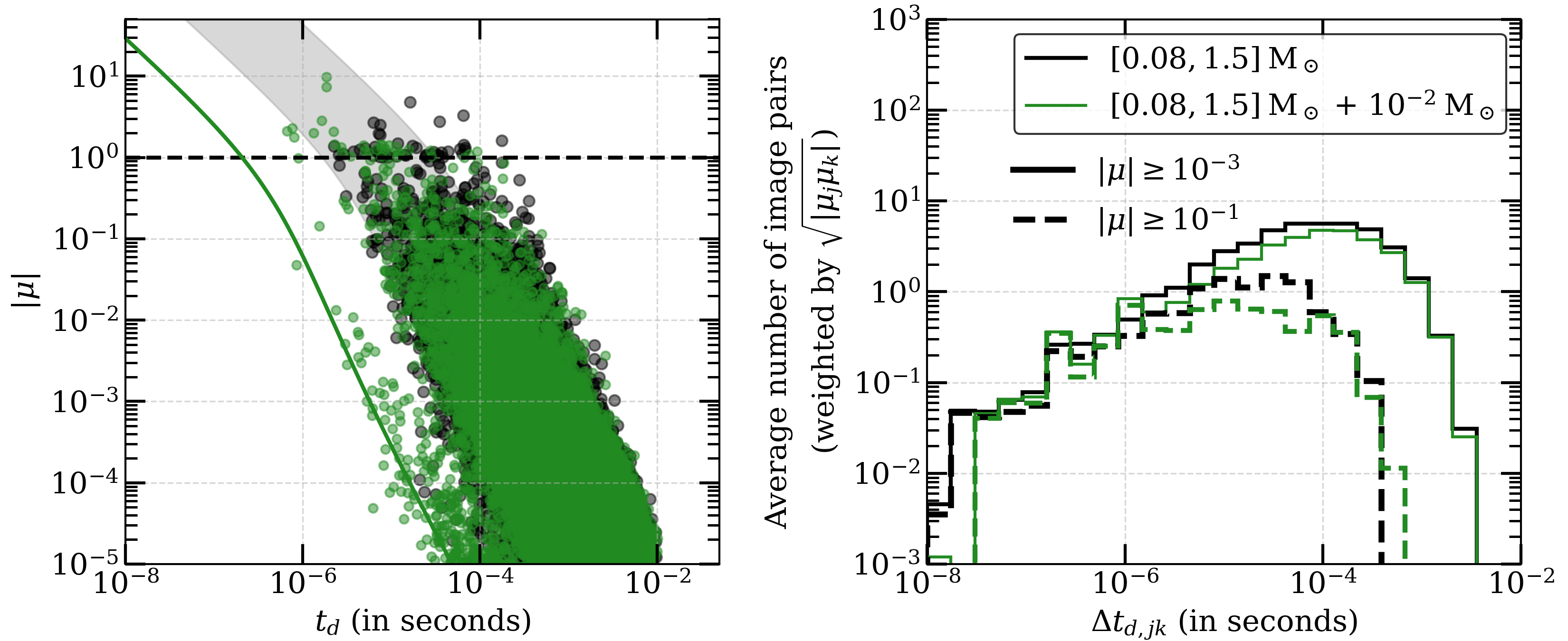}
    \caption{\textit{Left column.} Time delay~($t_d$) vs. absolute magnification~($|\mu|$) distribution of micro-images for a macro minima image with high optical depth. The black points represent the micro-images assuming a Salpeter mass function in~$[0.08, 1.5]~{\rm M_\odot}$ mass range. Green points are micro-images corresponding to a microlens population where 1\% of total convergence is in the form of~$10^{-2}\:{\rm M_\odot}$ point masses. The gray shaded region shows the region covering the range of~$|\mu|-t_d$ for an isolated point mass range of~$[0.08, 1.5]~{\rm M_\odot}$. The green curve corresponds to an isolated point mass of~$10^{-2}\:{\rm M_\odot}$. The horizontal black dashed line corresponds to~$|\mu|=1$. \textit{Right column.} Average histogram of time delays~($\Delta t_{d,jk}$) between pairs of micro-images shown in the left panel, weighted by the corresponding geometric mean of magnifications~$\sqrt{|\mu_j \mu_k|}$. The solid and dashed histograms include micro-images with~$|\mu|\geq10^{-3}$ and~$|\mu|\geq10^{-1}$, respectively.}
    \label{fig:high_tau}
\end{figure*}

Above, we saw that the stellar microlens density affects the time delay range of micro-image pairs. For Image-2 and Image-3, stellar microlenses alone can produce micro-image pairs with time delays comparable to micro-image pairs corresponding to the secondary microlens population in~$[10^{-3}, 10^{-2}]\:{\rm M_\odot}$, complicating the identification of distinct microlens populations via autocorrelation peaks. In contrast, Image-1, due to its lower stellar density, shows a clear distinction of microlens populations in micro-image pair time delays. Thus, the global minimum image is likely the most effective for detecting compact dark objects with fractions at the percent level. The average number of image pairs~(i.e., histogram height) for the stellar microlens case with~$|\mu|\geq0.1$ lies between 0.1 and 1 for all three images, implying that the average number of bright images~($|\mu|\geq1$) is less than one (except the brightest image). This may introduce challenges in identifying microlensing in a single strongly lensed FRB, as the smaller magnification will lead to lower signal-to-noise~(SNR) of autocorrelation peaks unless the SNR of the signal itself is very high. However, we may be able to mitigate this issue by using a repeating lensed FRB. For example, with stacking multiple bursts of a repeating lensed FRB with a repetition period of a few days or smaller, we can increase the possibility of detecting faint secondary micro-images, assuming that intrinsic variability from burst to burst is not significant. A repetition period of a few days makes sure that microlens-induced changes in micro-image time delays and magnifications are negligible. In a repeating lensed FRB with a large repetition period, we may encounter a scenario where the source crosses a micro-caustic in between two consecutive bursts, creating/destroying a pair of bright micro-images and significantly affecting the peak distribution in the autocorrelation. As the peaks in autocorrelation only depend on the time delay between pairs of micro-images and their magnification, we may not even need the lensed FRB to be repeating, and stacking multiple lensed FRBs with single bursts may also do the trick. We discuss this further in \Sref{sec:limitations} in more detail.

\section{Stellar mass function}
\label{sec:prob_imf}
In the above sections, we primarily focused on discriminating microlens populations with more than an order of magnitude difference in their average masses, which is relevant for differentiating stellar microlenses from dark compact objects. Another very interesting and important case is distinguishing different mass functions within the same mass range. For example, so far, we have assumed that the stellar population follows the Salpeter IMF. However, it is possible that galaxies may have a different (universal) IMF than the Salpeter one, or the IMF may vary across (and within) galaxies~\citep[e.g.,][]{2012ApJ...760...71C, 2015MNRAS.447.1033M}.

In this section, we ask: can we differentiate between various stellar mass functions given that we have the ability to resolve individual micro-images? To do so, we simulate stellar microlens populations for Image-1 and Image-2 from \Tref{tab:sl_system} using the Salpeter~\cite{1955ApJ...121..161S} and Kroupa~\cite{2001MNRAS.322..231K} mass functions. The results are shown in \Fref{fig:pm_sim_imf}. As expected, in the left column, all micro-images (black/blue points for Salpeter/Kroupa) closely follow the gray band, which represents the $|\mu|-t_d$ region for the saddle-point image corresponding to an isolated point mass lens. Although black and blue points show considerable overlap in the~$|\mu|-t_d$ plane, in the corresponding micro-image pair time delay histograms in the right column, we observe distinguishable features. 

For Image-1, due to the smaller stellar density, both Salpeter and Kroupa mass functions lead to very similar time delays between micro-image pairs. Although we do see an isolated peak in the histogram for the Kroupa mass function at~$\Delta t_{d,jk}\sim10^{-7}$~seconds, this seems to be an isolated event. However, as we focus on Image-2, the differences between the two mass functions become clear. The Salpeter mass function leads to more micro-images than the Kroupa mass function. In addition, the range of image pair time delays is also wider for the Salpeter mass function. Both of these differences can be attributed to the shape of these mass functions. The Salpeter mass function is bottom-heavy, whereas Kroupa is relatively flat at the low mass end. Due to that, the Salpeter mass function leads to more microlenses with smaller masses and gives rise to more micro-images than the Kroupa mass function. This is also reflected in the corresponding average mass values, which are 0.20~${\rm M_\odot}$ and 0.33~${\rm M_\odot}$ in the $[0.08, 1.5]~{\rm M_\odot}$ mass range for Salpeter and Kroupa mass functions, respectively. Contrary to the previous section, differences between the two stellar mass functions become more highlighted for Image-2 due to the higher stellar density compared to Image-1, implying that higher stellar densities are more preferred for differentiating between stellar mass functions.

\section{High optical depth}
\label{sec:high_tau}
In low microlensing optical depth, typically one micro-image dominates the total magnification, as we saw in previous sections. As we move towards high optical depths~($\tau\gtrsim0.5$), we can encounter multiple bright micro-images~($|\mu|\geq1$) primarily due to the formation of new micro-minima as discussed in~\citet{2011MNRAS.411.1671S}, making it relatively easy to detect micro-images in the lensed signal. However, an increase in the number of microlenses also brings the question of whether we still observe multiple branches of micro-images in the~$|\mu|-t_d$ plots due to different microlens mass functions and how the peak distribution in autocorrelation changes. Hence, in this section, we study the micro-image formations in high optical depth. Following~\citet{2020MNRAS.497.1583L}, we use $(\kappa, \gamma) = (0.5, 0.0)$ for a macro minimum image. Keeping in mind the extreme optical depth (and large number of microlenses), we only simulate two cases:
\begin{enumerate}[(i)]
    \item 100\% of total convergence is in the form of stellar microlens population drawn using the Salpeter mass function,
    \item 99\% of total convergence is in the form of stellar population with Salpeter mass function and 1\% in the form of microlenses with~$M=10^{-2}\:{\rm M_\odot}$.
\end{enumerate}

The results are shown in \Fref{fig:high_tau}. In the left panel, looking at black points, we note two obvious features that are the formation of many bright micro-images and micro-images leftwards of the gray shaded region, both of which can be explained by the formation of additional micro-minima. Minima images cannot be demagnified, implying the presence of bright images. Similarly, minima form valleys in the arrival time delay surface, meaning that such images will have lower time-delay values compared to the saddle points. Hence, they do not need to follow the gray shaded region, which corresponds to saddle point images formed due to isolated point masses in~$[0.08, 1.5]\:{\rm M_\odot}$ range, and can lie on the left of it. This does not mean that all micro-images leftward of the gray shaded region are micro-minima. For example, as shown in Fig.~1 of~\citet{2011MNRAS.411.1671S}, the presence of a microlens close to a minima can further divide it into two minima and one saddle point, and this saddle point can lie leftwards of the gray shaded region as it is forming in a valley in the arrival time delay surface. That said, even now, all of the stellar microlens-led micro-images still cluster around the gray shaded region. 

Green points in \Fref{fig:high_tau} correspond to the case~(ii). For~$|\mu|\lesssim0.1$, similar to previous sections, we see micro-images forming around the green curve. On the other hand, for~$|\mu|\geq0.1$, we do see additional micro-images but close to (and within) the gray shaded region rather than the green curve. This seems to imply that, at high optical depths, the presence of stellar microlenses close to micro-images formed due to secondary population significantly affects their properties, which can pose a challenge for discriminating between different stellar and $10^{-2}\:{\rm M_\odot}$ microlens populations. This becomes even more apparent as we look at the micro-image pair time delay distributions, as the corresponding histograms in black and green nearly follow each other. This, as well as the results for Image-2 and Image-3, can be explained by the fact that the fraction of compact dark matter is much smaller compared to stellar optical depth, and the micro-image formation is dominated by the latter. Hence, based on this, we can say that larger stellar optical depths increase the chance of detecting microlensing (as well as discriminating different stellar mass functions), but not for finding dark compact object population in the~$[10^{-3}, 10^{-2}]\:{\rm M_\odot}$ range on a percent level.

\section{Uncertainties}
\label{sec:limitations}
In the above sections, we saw that micro-images corresponding to different mass functions occupy distinguishable parts of the corresponding~$|\mu|-t_d$ plane. The resulting distribution of peaks in the autocorrelation, depending on the specific case, can enable us to probe the underlying microlens mass function(s). However, we did not consider any uncertainties or limitations coming from our lack of knowledge about the lens properties or the signal properties that can affect the significance of the detection of lensing features. In this section, assuming FRB as our source, we discuss possible uncertainties and their impact on the detectability and interpretation of microlensing signatures in the observed signal.

\subsection{Magnification \& SNR}
\label{ssec:limit_musnr}
FRBs are transient signals, such that for a typical case of FRB lensing by a galaxy, we observe only one macro-image at a time, as the time delay between macro-images can range from days to months. As discussed in \citet{2025arXiv250410523S}, thanks to micro-images, even when detecting only one counterpart of a strongly lensed FRB, we may be able to determine whether the underlying signal is lensed. However, without identifying the lensing galaxy and multiple counterparts of the lensed FRB, we cannot accurately model the strong lens mass distribution, leading to uncertainties in the corresponding macro-magnification estimates. Since the macro-magnification changes the effective microlens mass (and the optical depth), the uncertainty in macro-magnification will affect the~$|\mu|-t_d$ distribution of micro-images. For example, with Image-1, we can differentiate the stellar and dark microlens population in~$[10^{-3}, 10^{-2}]\:{\rm M_\odot}$ range as the latter leads to peaks in autocorrelation (at~$\Delta t_{j,k}<10^{-6}$~seconds) where the former cannot. However, an increase in the Image-1 macro-magnification is equivalent to increasing the corresponding optical depth and making micro-images brighter, in which case we can have a pair of stellar micro-images with similar delay as the micro-images formed due to dark compact objects, as we see for Image-2. Since the uncertainty in macro-magnification also affects the number of relevant micro-images, it would also introduce uncertainty in inferences regarding the stellar initial mass function. Larger error bars on the macro-magnification are typically expected for macro-images lying close to the critical curve. Hence, selecting a strong lens leading to macro-images with typical macro-magnifications, as done in \Sref{sec:sl_ml_pop}, can reduce these uncertainties.

The magnitude of peaks and the corresponding SNR values in the autocorrelation depend on the micro-image magnifications. In microlensing of a typical macro-minimum, on average, we expect less than one bright micro-image beyond the first micro-image. This raises the question of whether the corresponding autocorrelation peak will have sufficient SNR for detection. As we can see from \Eref{eq:auto}, the peaks in autocorrelation depend on the geometric mean of magnifications of pairs of micro-images, i.e.,~$C(t)\propto\sqrt{|\mu_j\mu_k|}$. Hence, demagnification in one of the micro-images can be compensated for by the other micro-image, allowing us to detect peaks corresponding to faint micro-images. Assuming that we have a repeating strongly lensed FRB, we can also look at the prospects for stacking multiple bursts. To do so, we need an FRB that is repeating on a period of days (or less) so that the microlens distribution remains the same. A more serious problem in repeating FRBs is variation in burst properties in time~\citep[e.g.,][]{2023ApJS..269...17H}. Although significant intrinsic variation from burst to burst in the signal is undesirable while searching for microlensing signatures, separating intrinsic variability from microlensing features will also help us reduce the error on strong lensing time delay measurements between different macro-images, as well as in the estimates of the Hubble constant~($H_0$). 

Another possible way to increase the significance of microlensing features in autocorrelation is to stack multiple different strongly lensed FRBs. Assuming that different microlens populations can lead to peaks in autocorrelation at different times, stacking different lensed FRBs may lead us to a bi-modal broad distribution corresponding to different populations. Here, one needs to carefully choose lensed FRBs with similar macro-magnification and preferably lower stellar microlens density, as it affects the time delay distribution of micro-images. Additional care, based on the properties of the observed signal, may be needed in deciding which lensed FRBs to stack, as stacking FRBs with drastically different properties may decrease the significance of microlensing detection. Given the large event rate, in the near future, we should be able to have a large number of lensed FRBs. A more detailed study is required in this direction to determine the feasibility of such an analysis, and it is subject to our future work.

\subsection{Microlens populations}
\label{ssec:limit_mlpop}
In our analysis, we have assumed that the stellar population in the lensing galaxy corresponds to individual microlenses in~$[0.08, 1.5]\:{\rm M_\odot}$ range with a given mass function such as Salpeter or Kroupa. However, at the time of birth, stars can have higher masses and at the end of their lives, stars with initial masses~$\gtrsim1.5\:{\rm M_\odot}$ turn into stellar remnants~(such as white dwarfs, neutron stars, and black holes) with masses~$\gtrsim0.4\:{\rm M_\odot}$~\citep[e.g.,][]{2017PASA...34...58E}. Since these remnants have masses in the stellar mass range, we do not expect them to give rise to micro-images having~$|\mu|-t_d$ similar to compact dark objects with masses~$[10^{-3}, 10^{-2}]\:{\rm M_\odot}$. However, we observe peaks in the autocorrelation, and their distribution can be affected by the increase in the number of micro-images due to the presence of stellar remnants polluting our inferences about the presence of different mass functions or the underlying stellar mass function. Hence, we also need to take into account the effects of such remnants on the micro-image distributions while making any predictions about the microlens (stellar or dark) mass function. 

Due to the choice of isolated stellar microlenses, we also neglected the effect of stellar multiplicity. A considerable fraction of stars are expected to be in binary systems~\citep[e.g.,][]{2013ARA&A..51..269D}. Based on the ratio of their orbital radii~($\theta_r$) and the corresponding Einstein angle~($\theta_{\rm E}$), we can divide binaries into three categories to model their lensing effects. Stellar binaries with~$\theta_r \ll \theta_{\rm E}$ can be treated as a point mass lens with an effective mass equal to the sum of component masses, whereas binaries with~$\theta_r \gg \theta_{\rm E}$ can be approximated as two isolated point mass lenses. Considerable deviations from an isolated point mass lens approximation can arise if~$\theta_r \sim \theta_{\rm E}$, where we can have additional micro-images with non-negligible magnification values. This tells us about how we can model the effect of binaries in our simulations. However, whether a stellar population in binaries will follow the same mass function as the isolated stars is another very important question. A more detailed study incorporating inputs from stellar modeling to assess the corresponding impact on the determination of the underlying microlens mass function is beyond the scope of the current work and left for the future.

\subsection{Plasma scattering}
\label{ssec:limit_plasma}
In the previous two subsections, we have discussed uncertainties primarily arising from our lack of knowledge about the macro- and micro-lens properties. Another uncertainty, maybe more severe, can arise from the scattering of the FRB signal caused by turbulent plasma present between the source and the observer~\citep[e.g.,][]{2021A&A...645A..44W}. The amount of scattering depends on the signal frequency~($f$) and the spatial gradient of underlying electron density. It can lead to deflections similar to microlensing and introduce new lensed images, resulting in additional peaks in the autocorrelation. From the arrival time delay surface perspective, turbulent plasma introduces an additional frequency-dependent term~($\propto f^{-2}$) in it, leading to new stationary points whose properties depend on the strength of density fluctuation in plasma and signal frequency~\citep[e.g.,][]{2023MNRAS.520..247K, 2025arXiv250410523S}. We note that such an effect will occur regardless of the macro-image type. If these new frequency-dependent peaks in the autocorrelation are present at every signal frequency, then we cannot segregate peaks arising from microlensing, and plasma scattering can introduce large errors in our inferences about the underlying microlens population. Since plasma scattering decreases at higher frequencies, we may expect cases with no autocorrelation peaks corresponding to plasma scattering at high frequencies. In such cases, we will be able to isolate microlensing peaks and decrease the uncertainty in our results.

Due to the extragalactic nature of the FRBs, the plasma scattering will have contributions from the host galaxy, lens galaxy, intergalactic medium, as well as the Milky Way. As we focus on galaxy-scale lenses, which are typically old with little to no star formation, they are not expected to have strong fluctuations (or turbulence) in the electron density, implying a low level of plasma scattering. That said, it is an open question as to what the lowest spatial scale is at which plasma inhomogeneities can exist and how strong they can be. The contribution from the intergalactic medium is also expected to be small due to low electron density and weak turbulence. Contribution from the Milky Way will depend on the line of sight, varying from microseconds at higher latitudes to milliseconds towards the galactic disk~\citep[e.g.,][]{2002astro.ph..7156C, 2017ApJ...835...29Y}. Thanks to the large FRB event rate, in the future, optimistically, we can only select lensed events at higher latitudes to limit the contribution from our own Galaxy. Keeping in mind that typical lensed sources are expected to be around $z\sim1-2$, the FRB host galaxy is very likely to be a star-forming galaxy. Hence, the contribution from the host galaxy (or in the vicinity of the FRB source) can be significant with scattering timescale reaching milliseconds~\citep[e.g.,][]{2021ApJS..257...59C, 2022ApJ...931...87O} and vary drastically from one FRB to another~\citep[e.g.,][]{2016ApJ...833..177S, 2017ApJ...834L...8M, 2024ApJ...963L..34G}. A more detailed study about the effects of plasma scattering on the inference of microlens mass function in strong lenses is subject to our future work and will be presented in upcoming papers.

Although in our current work, we have primarily focused on galaxy lenses, the same analysis is equally valid for galaxy cluster lenses. The two main differences between galaxy and cluster lenses are the microlens densities and the time delay between macro-images. In cluster lenses, the microlens density at the position of macro-images is typically expected to be~$\sim[1,50]~{\rm M_\odot\:pc^{-2}}$ range, which is smaller than galaxy lenses. This implies that for typical macro-magnification of~$\sim[1,10]$, we are in the low optical depth regime. Hence, thanks to the low stellar surface density, we may be able to probe a smaller fraction of compact dark matter than in the galaxy lenses as we saw in \Sref{sec:pml_pop} and for Image-1 in \Sref{sec:sl_ml_pop}. However, at the same time, plasma scattering is expected to be more significant (compared to typical galaxy lenses) due to higher turbulence in the intracluster medium~\citep[e.g.,][]{2023ApJ...949L..26C}.

\section{Conclusions}
\label{sec:conclusions}
FRBs, due to their high event rate, are great targets for gravitational lensing studies. Thanks to their point source nature and the outstanding time resolution (of the order of nanoseconds) achievable in the autocorrelation, they present us with excellent opportunities to detect effects of individual micro-images (more specifically, pairs of micro-images) in strongly lensed FRB signals. Observing such effects opens up the possibility to constrain the underlying microlens mass function within strong lenses. In our current work, primarily focusing on typical galaxy-scale lenses, we studied~$|\mu|-t_d$ distribution of micro-images corresponding to different microlens mass functions and the corresponding features in the autocorrelation. We found that low stellar microlens densities, similar to what we have close to the global minimum images, are preferred to probe a secondary microlens population on a percent level, having mass values in~$[10^{-3}, 10^{-2}]\:{\rm M_\odot}$ range. This becomes possible as the secondary population leads to peaks in the autocorrelation at times~($<10^{-6}$~seconds) where the stellar microlens population cannot. As we move towards higher stellar densities, due to the large number of micro-images, peaks from the secondary population become hard to segregate in the autocorrelation. That said, the chances of detecting stellar microlensing itself as well as distinguishing different stellar IMFs increase with an increase in stellar density. 

In (traditional) quasar microlensing, we observe a change in the observed brightness due to the collective microlensing, which is primarily sensitive to the average microlens mass~\citep[e.g.,][]{2001MNRAS.320...21W, 2020ApJ...904..176E}. In contrast, for FRB microlensing, we observe peaks stemming from pairs of micro-images in the autocorrelation, allowing us to gain further insights into the underlying microlens population(s). That said, for lensed FRBs, due to their transient nature, we cannot have long-term monitoring campaigns to better differentiate intrinsic and microlensing-induced features similar to lensed quasars~\citep[e.g.,][]{2022A&A...668A..77M}. Even with repeating ones, such campaigns appear challenging due to significant intrinsic variation from burst to burst and irregular intervals between bursts.

In our current work, we have mainly focused on lensing-induced features in the autocorrelation corresponding to different microlensing populations. However, the detectability of these features will be subject to various uncertainties coming from our lack of knowledge about the lens system or propagation effects. Although we have qualitatively discussed primary sources of uncertainties, more detailed studies are required to have a clear understanding and are subject to ongoing and future analyses. We also note that since the stellar population in lensing galaxies~(or in the ICM) is generally old, we will only be able to constrain the lower mass end of the stellar IMF. On the other hand, lensing of individual stars at cosmological distances \citep[known as caustic-crossing events; e.g.,][]{2018NatAs...2..334K} provides constraints on the high mass end of the stellar population in lensed galaxies~\citep[e.g.,][]{2025A&A...699A.299M}. With a combination of these different regimes, ultimately, we may be able to constrain the full mass range of the stellar IMF in the future.

\section*{Acknowledgements}
The authors thank Calvin Leung and Sung Kei Li for their useful comments. We also thank the anonymous referee for their useful comments and suggestions. This research has made use of NASA’s Astrophysics Data System Bibliographic Services. This work utilizes the following software packages: \textsc{python}~(\url{https://www.python.org/}), \textsc{matplotlib}~\citep{2007CSE.....9...90H}, \textsc{NumPy}~\citep{2020Natur.585..357H}, and \textsc{julia}~\citep[][\url{https://julialang.org/}]{bezanson2017julia}.

\noindent
\textit{Data Availability.} The code to generate data for this article is available in the Github repository: \url{https://github.com/akmeena766/GL_FRB_1}

\bibliographystyle{apsrev4-2}
\bibliography{Reference}

\end{document}